\documentclass[a4paper,12pt]{article}
\thicklines
\title{On Some rare weak decays of vector mesons.}
\author{L.M.~Kurdadze, Z.K.~Silagadze \vspace*{3mm} \\
\small \em
Budker Institute of Nuclear Physics \\
\small \em 630 090, Novosibirsk, Russia  \vspace*{3mm} \\}
\date{}

\begin{document}

\maketitle

\begin{abstract}
Some semileptonic weak decays of vector mesons are considered in the
framework of the most popular quark models. The predicted branching
ratios are unfortunately too small to make a study of these decays
realistic at meson factories under construction. 
\end{abstract}

\vspace{0.4cm}

\section{Introduction}
Weak decays of hadrons play an important role  in our understanding of
both perturbative  and nonperturbative aspects of  Standard Model.  On
the one hand they involve Kobayashi-Maskawa matrix elements and higher
order  corrections   to  weak  currents.   The latter  are  calculable
perturbatively to high  accuracy within the  Standard Model framework,
and the former are crucial parameters of the theory, not determined by
it, but  extracted   from   experiments. on  the    contrary,  another
ingredient of these weak decays, hadronic  matrix elements of the weak
currents are  not calculable at present from  the first principles and
are subject of nonperturbative QCD, the acronym which in reality means
a paradise for various phenomenological models of hadron structure. 

Semileptonic decays  with  $0^- \rightarrow 0^-$ and  $0^- \rightarrow
1^-$ hadron transitions attracted considerable  attention, as they are
promising  experimental sources  for extracting the  Kobayashi-Maskawa
matrix elements.  The  reviews of the  theoretical models, involved in
such a type  of exercise, along with  relevant literature can be found
in [1-4] and we  don't  repeat them  here.  Instead we 
focus  our efforts   on giving a  reliable estimate  for 
semileptonic decays with
$1^-   \to  0^-$   hadron  transitions.  Such   weak  decays   escaped
consideration   simply  because very   tiny  rates   are expected  for
them. Indeed, rough estimate of the semileptonic  decays rate is given
by the one  third  of the free  quark  decay width, assuming  that the
spectator antiquark is  irrelevant. It is  straightforward to get this
decay width \cite{5} 
       \begin{equation}            
         \Gamma(Q\rightarrow qe
         \bar{\nu})=\frac{G^2_F m^5_Q}{192\pi^3}|V_{qQ}|^2
         F(\frac{m_q}{m_Q}) \; , 
        \end{equation} 
\noindent
where $V_{qQ}$  is the  relevant Kobayashi-Maskawa matrix  element and
$F(x)=1-8x^2+8x^6-x^8-24x^4  \ln{x}$ is a   phase space factor, lepton
mass being neglected. 

Branching ratio, which follows from (1), for example, for $J/ \Psi \to
D^-_s e^+ \nu $ decay is about $10^{-9}$,  and other vector meson weak
semileptonic decay branching ratios appear to be even smaller. 

Below  we  use  two the most   popular  models to give  more elaborate
estimates for  the    vector meson semileptonic    decay  rates. These
calculations  were motivated by  the fact that several high luminosity
meson factories are expected to come into operation in near future. 

\section{General Considerations}

Let us   consider   $V(Q  \bar{Q})  \to    P(q\bar{Q})e^-   \bar{\nu}$
semileptonic  decay, where $V(Q  \bar{Q})$ and $P(q\bar{Q})$ stand for
vector $1^-$ and  pseudoscalar $0^-$ mesons,  made up from $Q \bar{Q}$
and  $q\bar{Q}$ quark-antiquark   pairs, respectively.   Corresponding
amplitude looks like 
$$A=\frac{G_F}{\sqrt{2}}~V_{qQ}~\bar{u}_e(\vec{k}_1)\gamma^\mu
(1-\gamma_5)v_\nu(\vec{k}_2)<P|J_\mu(0)|V> $$
\noindent
and  we will have  after averaging over  the vector meson polarization
and  summing over  the  leptons  spins  (lepton mass is  neglected and
$u^+u=2E$ normalization is used for lepton spinors) 
       \begin{equation}
        \overline{|A|^2}=\frac{1}{3}G_F^2|V_{qQ}|^2Sp(1-\gamma_5)
         \hat k_1\gamma^\mu\hat k_2\gamma^\nu \sum\limits_{s_v}
         <V|J_\nu^+(0)|P><P|J_\mu(0)|V> \; .
        \end{equation}
\noindent
Let us decompose \cite{6}
$$       \sum\limits_{s_V}<V(P,\epsilon)|J_\nu^+(0)|P(P^\prime)>
         <P(P^\prime)|J_\mu(0)|V(P,\epsilon)>= 
$$
        \begin{equation} \vspace*{-3mm}
         = -\alpha g_{\mu \nu}+
         \sum\limits_{\sigma_1,\sigma_2=\pm}\beta_{\sigma_1\sigma_2}
         (P+\sigma_1P^\prime)_\mu (P+\sigma_2P^\prime)_\nu+
         i\gamma\epsilon_{\mu \nu \lambda \sigma} (P+P^\prime)^\lambda
         (P-P^\prime)^\sigma  
        \end{equation} 

\vspace*{2mm} \noindent
and note   that terms from   (3) containing  $(P-P^{\prime})_{\mu}$ or
$(P-P^{\prime})_{\nu}$ don't contribute to (2), because, for example 
$$(P-P^{\prime})_{\mu}Sp(1-\gamma_5)  \hat  k_1\gamma^\mu\hat
k_2\gamma^\nu=  Sp(1-\gamma_5)\hat k_1(\hat k_1+\hat k_2)\hat
k_2 \gamma^\nu=0 $$ 
as lepton mass is assumed to be zero. 

So  only   $\alpha,\beta_{++}$   and  $\gamma$ invariant  form-factors
contribute to $\overline{|A|^2}$ and it is straightforward to get the
following expression   for  the  differential  width \cite{6,7} (in
$\vec{P}=0$ vector meson rest frame) 
$$      \frac{d^2\Gamma(V \to Pe\bar \nu)}{dxdy}= $$
        \begin{equation} \vspace*{-3mm}
        \frac{1}{3}~
        \frac{G_F^2M_V^5}{32\pi^3}|V_{qQ}|^2 \left \{ \alpha
        \frac{y}{M_V^2}+2\beta_{++}[4x(x_+-x)-y(1-2x)]-2\gamma
        y(x_+-2x+\frac{1}{2}y) \right \} \; ,   
        \end{equation}  

\vspace*{3mm} \noindent 
where $x_+=\frac{1}{2}(1-\frac{M^2_P}{M^2_V})$  and we have introduced
dimensionless  variables   $x=E_e/M_V$    and  $y=\frac{(P-P^{\prime})
^2}{M^2_V} \equiv\frac{t}{M^2_V}$, $E_e$ being the lepton energy. 

Thus decay width
       \begin{equation}
        \Gamma(V \to Pe\bar \nu)=\int \limits_{x_-}^{x_+}  
         \int \limits_{y_-}^{y_+} \frac{d^2\Gamma}{dxdy} \; , 
        \end{equation} 
\noindent with ($x_+$ was given above) \cite{7} 
$$x_-=0 \; , \: y_-=0 \; , \: y_+=\frac{4x(x_+-x)}{1-2x} \; .$$ 
\noindent
These integration limits are determined by decay kinematics. 

Note that for decays to $e^+ \nu$ the sign  of the term proportional to
$\gamma$ in (4) should  be reversed. The  simplest way to see  this is
the  following.   If lepton mass  is  neglected, when $\frac{d^2\Gamma(V 
\to P^-e^+ \nu)}{dxdy}$ can be obtained from $\frac{d^2\Gamma(V \to P^+
e^-\bar{\nu})}{dxdy}$ by replacement $x \to x^*=\frac{E_\nu}{M_V}$.  
But it    is   easy to  see   that   $x^*=x_+-x+\frac{1}{2}y$ and so  
$4x^*(x_+-x^*)-y(1-2x^*)= 4x(x_+-x)-y(1-2x)$,
but $x_+ -2x^* +\frac{1}{2}y=-(x_+ -2x+\frac{1}{2}y)$. 

It is convenient to  introduce form-factors which characterize hadronic
matrix element itself 
       \begin{eqnarray} &
        <P(P^\prime)|J_\mu(0)|V(P,\epsilon)>= & \nonumber \\ & = 
        ig\epsilon_{\mu \nu \lambda \sigma}\epsilon^\nu
        (P+P^\prime)^\lambda (P-P^\prime)^\sigma -f\epsilon_\mu
        -(\epsilon \cdot P^\prime)a_+(P+P^\prime)_\mu-
        (\epsilon \cdot P^\prime)a_-(P-P^\prime)_\mu \; . &    
        \end{eqnarray}
Comparing (6) and (3), and using
$$ \sum_{s_V} \epsilon_\mu \epsilon^*_\nu=
-g_{\mu \nu}+\frac{P_\mu P_\nu}{M^2_V},$$
\noindent
it is easy to find
       \begin{eqnarray} &
       \alpha=f^2+4M^2_V g^2 \vec {P}\,'^2 \; \; , \; \; \gamma=2gf
       \nonumber \\ & 
       \beta_{++}=\frac{f^2}{4M_V^2}-M_V^2g^2y+\frac{1}{2}
       \left [ \frac{M_P^2}{M_V^2}-y-1 \right ]fa_++a_+^2\vec {P}\,'^2  
       \; , &
       \end{eqnarray}
where
     $$ \vec{P}\,'^2=\frac{[M^2_V (1-y)+M^2_P]^2}{4 M^2_V}-M^2_P \; .$$

Another popular set of form-factors is defined through \cite{8}
        $$
        <P(P^\prime)|J_\mu(0)|V(P,\epsilon)>= \frac{2i}{M_V+M_P}
        \epsilon_{\mu \nu \lambda \sigma}\epsilon^\nu P^{\prime
         \lambda}P^\sigma~V(q^2)- $$  
         \begin{equation} \vspace*{-3mm}
         -\epsilon_\mu(M_V+M_P)~A_1(q^2)-\frac{\epsilon \cdot q}
         {M_V+M_P}(P+P^\prime)_\mu~A_2(q^2)+\cdots \; \; ,
         \; q=P-P^\prime \; . 
         \end{equation}

\vspace*{3mm}
Dots here is for terms proportional to $(P-P^\prime)_\mu$, which
do not contribute to decay width for massless lepton.

Obvious relations between these two sets of form-factors are
       \begin{equation}
        g(q^2)=\frac{V(q^2)}{M_V+M_P},\; \; f(q^2)=(M_V+M_P) A_1(q^2),
        \; \; a_+(q^2)=-\frac{A_2(q^2)}{M_V+M_P}.
        \end{equation}

Some model for hadron structure is needed to concretize the introduced
form-factors.

\section{ISGW model}          

The Isgur-Scora-Grinstein-Wise model \cite{9} uses nonrelativistic
quark model wave functions to predict weak hadronic form-factors.
Strictly speaking, this model becomes rigorous in the weak binding
limit where $M_V\approx 2m_Q$ and $M_P\approx m_Q + m_q$, and near
the zero recoil point where $t=q^2$ reaches its maximum value $t_m=
(M_V-M_P)^2$. But it is assumed that the resulting form-factor formulas
are valid even beyond the weak binding regime. More serious problem is
that the nonrelativistic quark model predictions about the $(t_m -t)$-
dependence of form-factors are not reliable when $t_m -t$ becomes large
compared to typical hadronic scales. Nevertheless this model proved
to be successful and up to now remains one the most popular one, maybe
because "it is better to have the right degrees of freedom moving at
the wrong speed than the wrong degrees of freedom moving at right
speed" \cite{10}. Updated version of the ISGW model, which incorporates
relativistic  corrections,heavy quark symmetry constraints and more
realistic behavior of form-factors at large $t_m -t$, is given in 
\cite{10}.

In the weak binding limit the state vectors of the nonrelativistic
$V(Q \bar Q)$ vector or $P(q \bar Q)$ pseudoscalar mesons can
be represented as a superposition of the free quark-antiquark states
\cite{9,11}
       \begin{eqnarray} & &
 |V(\vec P,\epsilon)>=\sqrt{2M_V}\int\frac{d\vec p}{(2\pi)^3}
\sum\limits_{ms\bar s}C^{1m}_{s\bar s}~\epsilon \cdot \epsilon^*_{(m)}
\varphi_{V}(\vec P)|Q[\frac{m_Q}{M_V}\vec P+\vec p,s]
\bar Q[\frac{m_Q}{M_V}\vec P-\vec p,\bar s]>, \nonumber \\ & &
        | P(\vec P)>=\sqrt{2M_P}\int\frac{d\vec p}{(2\pi)^3}
\sum\limits_{s\bar s}C^{00}_{s\bar s}
\varphi_{P}(\vec P)|q[\frac{m_q}{M_P}\vec P+\vec p,s]
\bar Q[\frac{m_Q}{M_P}\vec P-\vec p,\bar s]>. 
        \end{eqnarray}
\noindent
We use $<\vec P\,^\prime|\vec P>=(2\pi)^3 2E\delta (\vec P\,'-\vec P)$ 
normalization for the meson state vectors while
$<\vec p\,^\prime|\vec p>=(2\pi)^3~\frac{E}{m}~\delta(\vec p\,'-\vec p)$
one for the state vectors of quark (or antiquark) with mass m. 
$\epsilon_{(-)}, \; \epsilon_{(0)}$ and $\epsilon_{(+)}$ are three 
independent polarization $4$-vectors for the vector mesons. 
$C^{jj_z}_{s\bar s}$ are the usual
Clebsh-Gordan coefficients that couple $s$ and $\bar s$ quark and
antiquark spins into the meson spin and polarization. At last,
$|Q[\vec p_1,s]\bar Q[\vec p_2,s]>=a^+_s (\vec p_1) b^+_{\bar s}
(\vec p_2)|0>$, $a^+$ and $b^+$ being the quark and antiquark creation
operators. Note that our normalization convention indicates the 
following anticommutation relations
       \begin{equation}
\{a_s(\vec p),a^+_{s'}(\vec p\,') \}=\{b_s(\vec p),b^+_{s'}
(\vec p\,') \}=(2\pi)^3\frac{E}{m}\delta (\vec p\,'-\vec p) \; .
        \end{equation}

To obtain the quark model weak transition matrix element, one should
replace $J_\mu (0)$ weak current in $<P(P')|J_\mu (0)|V(P,\epsilon )>$
by the quark weak current $j_\mu (0)=\bar q(0)\gamma_\mu (1-\gamma_5)
Q(0) $  (the Kobayashi-Maskawa matrix element was already separated),
decompose quark field operator (note $u^{+(\lambda)}(\vec k)
u^{(\lambda^\prime)}(\vec k)=v^{+(\lambda)}(\vec k)v^{(\lambda^\prime)}
(\vec k)=\frac{k_0}{m} \delta_{\lambda \lambda ^\prime}$ normalization 
for the Dirac spinors)
$$\Psi(0)=\sum\limits_{\lambda}\int \frac{d\vec k}{(2\pi)^3}\frac{m}{k_0}
[a_\lambda (\vec k)u^{(\lambda)}(\vec k)+b^+_\lambda (\vec k)
v^{(\lambda)}(\vec k)] $$
\noindent
and use the anticommutation relations (11) along with the 
nonrelativistic approximation $E=\sqrt{m^2 +\vec p ^2}\approx m$. 
As a result we obtain
(in the vector meson rest frame $\vec P=0$)
$$       <P(P')|J_{\mu}(0)|V(P,\epsilon)>=\sqrt{4M_P M_V}
\int \frac{d\vec p}{(2\pi)^3} \varphi^*_P (\frac{m_Q}{M_P} 
\vec{P}'+\vec{p})\varphi_V (\vec p) \times $$
\begin{equation} \hspace*{-3mm}
\times \sum\limits_{mss'\bar s} C^{00}_{s'\bar s}
C^{1m}_{s\bar s} \epsilon \cdot \epsilon^*_{(m)}
\bar u^{(s')}_{(q)}(\vec P\,'+\vec p)\gamma_\mu (1-\gamma_5)
u^{(s)}_{(Q)}(\vec p) . 
        \end{equation}
\noindent
To simplify (12), note that
$$u^{(s)}(\vec p)=\frac{(\hat p+m)}{\sqrt{2m(p_0+m)}} \chi^{(s)}
\approx \frac{(\hat p+m)}{2m} \chi^{(s)} \: , \; p_0\approx m, $$
\noindent
where $\chi^{(s)}$ is the rest frame spinor, and
          \begin{eqnarray} &
\sum\limits_{mss'\bar s} C^{00}_{s'\bar s} ~\epsilon \cdot 
\epsilon^*_{(m)}
C^{1m}_{s\bar s} \chi^{(s)} \bar{\chi}^{(s')}=
\frac{1}{2} \{(\chi^{(+)}\bar{\chi}^{(+)}-\chi^{(-)}\bar{\chi}^{(-)})
\epsilon \cdot \epsilon^*_{(0)}-\sqrt{2} \chi^{(+)}\bar{\chi}^{(-)} 
\epsilon \cdot \epsilon^*_{(+)}+ & \nonumber \\ & +
\sqrt{2} \chi^{(-)}\bar{\chi}^{(+)} \epsilon \cdot \epsilon^*_{(-)} \}= 
\frac{1}{4}(1+\gamma_0) \{\gamma_3 \epsilon \cdot \epsilon^*_{(0)} +
\gamma_+ \epsilon \cdot \epsilon^*_{(+)}+\gamma_- \epsilon \cdot 
\epsilon^*_{(-)}\}\gamma_5= & \nonumber \\ &=\frac{1}{4}(1+\gamma_0 ) 
\vec{\gamma} \cdot \vec{\epsilon} \gamma_5 . & \nonumber
           \end{eqnarray}
\noindent
Here
$\gamma_+=-\frac{1}{\sqrt2}(\gamma_1 +i \gamma_2), \;
\gamma_-=\frac{1}{\sqrt2}(\gamma_1 -i \gamma_2)$ and 
$\epsilon^*_{\pm}=-\epsilon_{\mp}$
property was used in the last step (note that
$\vec \gamma \cdot \vec \epsilon=\sum\limits_{s=0,\pm}(-1)^s 
\gamma_{-s}\epsilon_{(s)}$).

Thus (12) transforms into
       $$
    <P(P')|J_{\mu}(0)|V(P,\epsilon)>=\frac{\sqrt{M_P M_V}}{8m_qm_Q}
\int \frac{d\vec p}{(2\pi)^3} \varphi^*_P (\frac{m_Q}{M_P} 
\vec{P}'+\vec{p}) \varphi_V (\vec p) \times $$
        \begin{equation} \vspace*{-3mm}
 Sp\{(1+\gamma_0 ) \vec{\gamma} \cdot \vec{\epsilon} \gamma_5
(\hat p'+m_q)\gamma_\mu (1-\gamma_5)(\hat p+m_Q)\}  , 
        \end{equation}

\vspace*{2mm}
where $p\,'_0=m_q$ and $\vec p\,'=\vec P\,'+\vec p$.

It is now straightforward to extract the Lorentz-invariant form-factors
from (13) once $\varphi_P$ and $\varphi_V$ wave functions are specified.
It is assumed in the ISGW approach that in the role of these wave 
functions one should use Schr\"{o}dinger wave functions for the usual 
Coulomb plus linear potential that proved to be useful in quarkonia 
spectroscopy. But to simplify
numerical calculations, they in fact used variational solutions of this
Schr\"{o}dinger problem based on Gaussian type harmonic-oscillator wave 
functions. In our case the relevant trial function is
$\varphi (\vec r)=\frac{\beta ^{3/2}}
{\pi ^{3/4}}\exp{(-\beta^2r^2/2)}$,
its momentum space image being
       \begin{equation}
   \varphi (\vec p)=\left(\frac{2\sqrt\pi}{\beta}\right)^{3/2} 
   \exp{(- \vec p\,^2/2\beta^2)}
        \end{equation}
with $\beta$ as variational parameter.

Let us introduce designations
$$<A_0>=\frac{1}{4}Sp\{(1+\gamma_0 ) \vec{\gamma} \cdot \vec{\epsilon} 
\gamma_5 (\hat p\,'+m_q)\gamma_0 \gamma_5(\hat p+m_Q)\} $$
and analogously for $<\vec A>,\; <V_0>$ and $<\vec V>$. When we will 
have in the nonrelativistic limit \cite{11}
      \begin{eqnarray} &&
<A_0>=(m_Q+p_0)\vec p\,'\cdot\vec\epsilon+(m_q+p\,'_0)\vec p\cdot 
\vec\epsilon \to
2\{m_Q(\vec P\,'+\vec p)\cdot\vec \epsilon +m_q \vec{p}\cdot
\vec{\epsilon} \}
\nonumber \\ &&
<\vec A>=\vec{\epsilon}\cdot\vec{p}\,'~\vec{p}+
\vec{\epsilon}\cdot\vec{p}~\vec{p}\,'+
p\,'\cdot p ~\vec{\epsilon}+m_q p_0 ~\vec{\epsilon}+m_Q p\,'_0 
~\vec{\epsilon}+m_q m_Q~\vec{\epsilon} 
\to \nonumber \\ && 
\to 4m_q m_Q\vec{\epsilon}-\vec{p}\cdot(\vec{P}\,'+\vec{p})
\vec{\epsilon}+
(\vec{p}\cdot\vec{\epsilon})(\vec{P}\,'+\vec{p})+(\vec{P}\,'\cdot
\vec{\epsilon})\vec{p}+(\vec{p}\cdot\vec{\epsilon})\vec{p} , \\ 
&&  <\vec V>=i\{(m_Q+p_0)\vec{\epsilon}\times\vec{p}\,'-
    (m_q+p'_0)\vec{\epsilon}\times\vec{p} \} \to  
    2i\{m_Q\vec{\epsilon}\times (\vec{P}\,'+\vec{p})-
    m_q\vec{\epsilon}\times \vec{p} \} \nonumber
        \end{eqnarray}
Using the last expression in (15) along with the equalities
       \begin{equation}
\int \frac{d\vec p}{(2\pi)^3} \varphi^*_P (\frac{m_Q}{M_P} 
\vec P\,'+\vec p)
\varphi_V (\vec p)=\left(\frac{\beta_P \beta_V}{\beta^2_{PV}}
\right)^{3/2}
\exp{\left \{-\frac{m^2_Q}{4M_P M_V}~\frac{t_m -t}{\beta^2_{PV}}
\right\}}\equiv F(t),
        \end{equation}
\noindent
where $\beta^2_{PV}=\frac{1}{2}(\beta^2_P+\beta^2_V)$, and
       \begin{equation}
\int \frac{d\vec p}{(2\pi)^3} \varphi^*_P (\frac{m_Q}{M_P} \vec P\,'
+\vec p) ~\vec p ~
\varphi_V (\vec p)= -\frac{m_Q}{M_P} ~\frac{\beta^2_V}{2\beta^2_{PV}}
F(t)\vec P\,^\prime \; ,
        \end{equation}
\noindent
we get from (13) (it is supposed that the vector weak current $\vec V(0)$ 
will be not confused with the vector meson $V$)
$$<P(P')|\vec {V}(0)|V(P,\epsilon)>=i \vec \epsilon \times \vec {P}\,'
\sqrt{M_P M_V}\left\{\frac{1}{m_q}-\frac{1}{2\mu_-}\frac{m_Q}{M_P}
\frac{\beta^2_V}{\beta^2_{PV}} \right\}F(t) \; ,$$
\noindent
where
       \begin{equation}
        \mu_{\pm}=\left[\frac{1}{m_q} \pm \frac{1}{m_Q}\right]^{-1} .
        \end{equation}

On the other hand according to (16) we should have in the $\vec P=0$
frame
$$<P|\vec V(0)|V>=2i M_V g \vec\epsilon \times \vec P \, '.$$

Comparing these two expressions, we immediately get
       \begin{equation}
        g=\frac{1}{2}\sqrt{\frac{M_P}{M_V}}F(t) \left \{\frac{1}{m_q}
        -\frac{1}{2\mu_-}\frac{m_Q}{M_P}
        \frac{\beta^2_V}{\beta^2_{PV}} \right \} \; .
        \end{equation}

Analogously the first equation in (15) leads to
       \begin{equation}
        a_+(M_P+M_V)+a_-(M_V-M_P)=-\sqrt{M_P M_V}\left\{\frac{1}{m_q}
-\frac{1}{2\mu_+}\frac{m_Q}{M_P}
\frac{\beta^2_V}{\beta^2_{PV}} \right\}F(t).
        \end{equation}

There is some subtlety in using the equation for $<\vec A>$ from (15).
For $\vec\epsilon ~\bot \vec P'$ polarization it readily gives
       \begin{equation}
        f=2\sqrt{M_P M_V}F(t),
        \end{equation}
\noindent
while for $\vec\epsilon ~\| \vec P'$ polarization it involves 
$\sim \vec p^2$
terms about which there is no guarantee in our nonrelativistic approach.
Nevertheless one can get the correct answer by separating $D$-wave 
partial amplitude, because there is nothing intrinsically relativistic 
in recoiling into a $D$ wave [9]. So let us disregard 
$\sim \vec \epsilon$ terms from
$<\vec A>,$ that correspond to a $S$-wave, and also in
       \begin{equation}
     \int \frac{d\vec p}{(2\pi)^3} \varphi^*_P ( \vec{p}+\vec {q})p_i p_j
      \varphi_V (\vec p)=A\vec q^2\delta_{ij}+B q_i q_j,
       \end{equation}
\noindent
omit the first term, which leads to $S$-wave amplitude too. Using
$$\int \frac{d\vec p}{(2\pi)^3} \varphi^*_P ( \vec{p}+\vec{q})\vec p\,^2
\varphi_V (\vec p)=\left[\frac{3}{2}\frac{\beta^2_P\beta^2_V}
{\beta^2_{PV}}
+\frac{q^2}{4}\frac{\beta^4_V}{\beta^4_{PV}} \right] F(t)$$
\noindent
and
$$\int \frac{d\vec p}{(2\pi)^3} \varphi^*_P ( \vec {p}+\vec {q})
(\vec p \cdot \vec q)^2 \varphi_V (\vec p)=
\left[\frac{1}{2}\frac{\beta^2_P\beta^2_V}{\beta^2_{PV}}+
\frac{q^2}{4}\frac{\beta^4_V}{\beta^4_{PV}} \right]\vec q\,^2 F(t),$$
\noindent
we easily obtain
       \begin{equation}
        B=\frac{1}{4}\frac{\beta^4_V}{\beta^4_{PV}}F(t).
        \end{equation}

Now we have all necessary ingredients to get a relation which follows 
from the $D$-wave relevant terms of $<\vec A>$:
       \begin{equation}
 a_+-a_-=\frac{\sqrt{M_PM_V}}{m_qm_Q}F(t)\left[\frac{m_Q}{M_P}
~\frac{\beta^2_V}{2\beta^2_{PV}}
-\frac{1}{4}\frac{m^2_Q}{M^2_P}~\frac{\beta^4_V}
{\beta^4_{PV}}\right] \; .            
        \end{equation}

From (20) and (24) $\; a_+$ form factor can be evaluated. Nothing that 
in the weak binding approximation $\frac{M_V-M_P}{m_qm_Q}
\approx \frac{m_Q-m_q}{m_qm_Q}=\frac{1}{\mu_-}$, we get
$$a_+=\sqrt{\frac{M_P}{M_V}}F(t)\left\{\frac{1}{2m_q}\frac{m_Q}{M_P}
~\frac{\beta^2_V}{\beta^2_{PV}}
-\frac{1}{8\mu_-}\frac{m^2_Q}{M^2_P}~\frac{\beta^4_V}{\beta^4_{PV}}-
\frac{1}{2m_q}\right\}\; .$$

Let us further transform
         \begin{eqnarray*}       
\frac{1}{m_q}~\frac{m_Q}{M_P}~\frac{\beta^2_V}{\beta^2_{PV}}=
\frac{1}{m_q}~\frac{m_Q}{M_P}~\frac{(\beta^2_V-\beta^2_P)+
(\beta^2_V+\beta^2_P)}{(\beta^2_V+\beta^2_P)}= \\ =
\frac{1}{m_q}~\frac{m_Q}{M_P}~\frac{\beta^2_V-\beta^2_P}
{\beta^2_V+\beta^2_P}+\frac{1}{m_q}~\frac{m_Q+m_q-m_q}{M_P}\approx
\frac{1}{m_q}-\frac{1}{M_P}+\frac{1}{M_P}~\frac{m_Q}{m_q}
~\frac{\beta^2_V-\beta^2_P}{\beta^2_V+\beta^2_P} \; . 
          \end{eqnarray*} 

This enables to rewrite $a_+$ form factor as
       \begin{equation}
a_+=\sqrt{\frac{M_P}{M_V}}~\frac{F(t)}{2M_P}\left[-1+\frac{m_Q}{m_q}
~\frac{\beta^2_V-\beta^2_P}{\beta^2_V+\beta^2_P}- 
\frac{1}{4\mu_-}~\frac{m^2_Q}{M_P}~\frac{\beta^4_V}{\beta^4_{PV}}
\right] \; .
        \end{equation}
                                        
Having at hand (19), (21) and (25) expressions for the $g,f$ and $a_+$
form factors, the semileptonic decay width can be evaluated through
(4), (5) and (7) formulas.

\section{BSW model}

The Bauer-Stech-Wirbel model [8,12] uses the quark model to deal only
with one point $q^2=0$. In contrast to the zero recoil point, considered
previously in the ISGW model, $q^2=0$ point can be highly relativistic.
So the relativistic treatment of quark dynamics becomes unavoidable, 
although this dynamics greatly simplifies in the Infinite Momentum Frame.
It is convenient to represent meson state vectors in this frame in 
the slightly different from (10) form
        \begin{eqnarray} \hspace*{-5mm} & & 
|V(P,\epsilon )>=\sqrt{2} \sum\limits_{s\bar s m}\int \frac
{d\vec p_1d\vec p_2}
{(2\pi)^{3/2}}\sqrt{\frac{m_Qm_Q}{p_{10}p_{20}}}
\delta (\vec P-\vec p_1-\vec p_2)C^{1m}_{s\bar s}\epsilon \cdot 
\epsilon^*_m
\varphi_V(\vec p_1)|Q[\vec p_1,s]\bar Q[\vec p_2,\bar s ]>, 
\nonumber \\ \hspace*{-5mm} & & 
|P(P')>=\sqrt{2} \sum_{s\bar s }\int \frac{d\vec p_1d\vec p_2}
{(2\pi)^{3/2}}\sqrt{\frac{m_qm_Q}{p_{10}p_{20}}}
\delta (\vec P\,'-\vec p_1-\vec p_2)C^{00}_{s\bar s} 
\varphi_P(\vec p_1)|q[\vec p_1,s]\bar Q[\vec p_2,\bar s ]>.        
        \end{eqnarray}
In the Infinite Momentum Frame and for $\vec q =\vec P_V-\vec P_P=0$
we have
$P_{V\mu}=(E_V,0,0,P),\;$ $ P_{P\mu}=(E_P,0,0,P),\;  P \to \infty$.
But $E_V-E_P\approx \frac{M^2_V-M^2_P}{2P} \to 0$. That is $\vec q=0$
just gives $q^2=0$ point.

Let us introduce the longitudinal momentum fraction carried by the active 
quark in the meson $x=\frac{p_{1z}}{P}$, when the normalization condition
for the $\varphi (\vec p_1)$ wave faction, which follows from (26), is
             \begin{equation}
        \int dx d\vec p_T |\varphi(x,\vec p_T)|^2 =1.
             \end{equation}
             
The concrete form of this wave function is inspired by the relativistic 
harmonic oscillator model and looks like \cite{8} (for the meson of mass 
$M$ made up from active 
quark $q$ and spectator antiquark $\bar Q$)
             \begin{equation}
      \varphi (x,\vec p_T)=N \sqrt{x(1-x)}~\exp{\left(-\frac{\vec p^2_T}
        {2w^2}\right)}~
     \exp{\left\{-\frac{M^2}{2w^2}\left(x-\frac{1}{2}-\frac{m^2_q-m^2_Q}
        {2M^2}\right)^2\right\}} \; ,
             \end{equation}             
\noindent             
where $N$ is determined from the normalization condition (27). 
The dimensional parameter $\omega$ controls transverse momentum 
suppression and equals to the average transverse momentum 
$\omega^2=<\vec p\,^2_T>$. 
In the role of $\omega$ we can use $\beta$ parameter from (14), as 
$\vec p_T$ is not changed by the boost along $z$-direction.

     Manipulations which had leaded to (12), now give for 
$\vec q=\vec P_V - \vec P_P=0$
       \begin{eqnarray} &
       <P(P)|J_{\mu}(0)|V(P,\epsilon)>= & \nonumber \\
&  2 \sum\limits_{mss'\bar s}
\int d\vec p \sqrt{\frac{m_qm_Q}{p_0 p'_0}}C^{00}_{s'\bar s}
C^{1m}_{s\bar s} \epsilon \cdot \epsilon^*_{(m)}
 \varphi^*_p (\vec p)\varphi_V (\vec p) 
\bar u^{(s')}_{(q)}(\vec p)\gamma_\mu (1-\gamma_5)
u^{(s)}_{(Q)}(\vec p) \; . & 
        \end{eqnarray}

In the infinite momentum limit $p_0=\sqrt{m^2_Q+\vec p^2}=
\sqrt{m^2_Q+x^2P^2+\vec p^2_T} \to xP\;$, $ \; p\,'_0=
\sqrt{m^2_q+\vec p^2} \to xP$ and
$$ u^{s}(\vec p)=\frac{\hat p+m}{\sqrt{2m(p_0+m)}} \chi^{(s)} \to
\frac{\hat p+m}{\sqrt{2mxP}}\chi^{(s)}. $$     
\noindent So (29) transforms into
             \begin{equation}
 <P|J_{\mu}(0)|V>=\int dx d\vec p_T \frac{\varphi^*_p (\vec p)\varphi_V 
(\vec p)}{x^2P} 
 Sp\{\frac{1}{4}(1+\gamma_0 ) \vec{\gamma} \cdot \vec{\epsilon} \gamma_5
(\hat p\,'+m_q)\gamma_{\mu} (1-\gamma_5)(\hat p+m_Q)\},            
             \end{equation} 
\noindent 
where $\vec p\,'=\vec p$ and $p\,'_0 \to xP$.

      From (15) we will have in the $P \to \infty$ limit
      
   $$<\vec V>\to i(m_Q-m_q)\vec \epsilon \times \vec p .$$

\noindent On the other hand according to (6)
$$<P|\vec V(0)|V>=2 i g (E_V-E_P) \vec \epsilon \times \vec P_V \to
i g \frac{M^2_V-M^2_P}{P} \vec \epsilon \times \vec P_V  .  $$

Comparing these expressions of $<P|\vec V(0)|V>$ and using
$$\int d \vec p \frac{1}{x^2 P}  \varphi^*_P (\vec p) \vec p \varphi_V 
(\vec p) \to \frac{1}{P} J \vec \epsilon \times \vec P_V , $$ 
\noindent
where
         \begin{equation}
J=\int d \vec p_T \int^1_0 \frac{dx}{x} \varphi^*_P (x,\vec p_T)
\varphi_V (x, \vec p_T) ,
         \end{equation}
\noindent we get 
        \begin{equation}
        g(q^2=0)=\frac{m_Q-m_q}{M^2_V-M^2_P} J .
        \end{equation}

But this gives the form factor only at one $q^2=0$ point. For values of 
$q^2$ others than zero, the BSW model assumes nearest pole dominance:
        \begin{equation}
A_1(q^2)=\frac{h_{A_1}}{1-\frac{q^2}{M^2_{1^+}}}, \; \;
A_2(q^2)=\frac{h_{A_2}}{1-\frac{q^2}{M^2_{1^+}}}, \; \;        
V_1(q^2)=\frac{h_V}{1-\frac{q^2}{M^2_{1^-}}}, 
        \end{equation}

Thus the $q^2$-dependence of form factors are determined once the masses 
of the appropriate $1^-$ and $1^+$ vector mesons are known.

Then (32) indicates, that
        \begin{equation}
        h_V=\frac{m_Q-m_q}{M_V-M_P} J .  
        \end{equation}

Analogously, using $<\vec A> \to x(m_Q+m_q)P ~\vec \epsilon + 
2(\vec\epsilon \cdot \vec p) ~\vec p , \; \; 
<P|\vec A(0)|V>=f\vec\epsilon+\linebreak 2(\epsilon \cdot P_P)a_+ 
\vec P_V$ and $\epsilon \cdot P_P \to 
(\frac{E_P}{E_V}-1) \vec \epsilon \cdot \vec P_V $, we can get 
$f=(M_Q+M_q)J$, 
\noindent and so
        \begin{equation}
        h_{A_1}=\frac{m_Q+m_q}{M_V+M_P} J .  
        \end{equation}

Again there is a subtlety in extracting $a_+$. Instead of giving a 
rigorous derivation, we prefer the following educative guess. 
Noting that for $P_{V\mu}=(E_V,0,0,P)$ the longitudinal
polarization 4-vector $\epsilon_{\|\mu}=\frac{1}{M_V}(P,0,0,E_V) \to 
\frac{P}{M_V}(1,0,0,1)$,
\noindent we obtain
$$\int d\vec x <P|A_0(x)|V>=(2\pi)^3 \delta (\vec P_V-\vec P_P)
<P|A_0(0)|V> 
\to $$ $$ \to (2\pi)^3 \delta (\vec P_V -\vec P_P)\left \{f \epsilon_0+
\left(\frac{E_P}{E_V}-1\right) \epsilon_3 P(E_V+E_P)a_+\right \} 
\to $$ $$ \to (2\pi)^3 \delta (\vec P_V -\vec P_P)\left\{f-(M^2_V-M^2_P)
a_+\right\}\frac{P}{M_V} .$$

On the other hand $Q_{50}=\int d\vec x A_0(x)$ is an appropriate weak 
charge, which in the  exact flavor symmetry limit transforms $|V>$ 
initial state into $|P>$ final state and so
$$\int d\vec x <P|A_0(x)|V>=<P(P_P)|Q_{50}|V(P_V)>=<P(P_P)|P(P_V)> 
\to  2P(2\pi)^3 \delta (\vec P_V-\vec P_P).$$

In the broken flavor symmetry case one should expect instead 
$<P(P_P)|Q_{50}|V(P_V)>=2PI(2\pi)^3 \delta (\vec P_V-\vec P_P)$, with 
$I$ as the wave function overlap integral.
        \begin{equation}
I=\int d\vec p_T \int^1_0dx \varphi^*_P (x,\vec p_T)
\varphi_V (x,\vec p_T) \; . 
        \end{equation}
\noindent Thus we obtain
    $$a_+=\frac{1}{M^2_V-M^2_P}[f-2M_VI]$$
and so
        \begin{equation}
   h_{A_2}=\frac{2M_V}{M_V-M_P} I - \frac{M_V+M_P}{M_V-M_P} h_{A_1}.  
        \end{equation}

(34), (35) and (36) formulas and the nearest pole dominance hypothesis 
completely determine the weak form factors in the BSW model.

\section{Heavy quark limit}

In the limit in which the quarks active in weak transition are very heavy, 
all form factors for this transition can be expressed in terms of a single 
function $\xi(\zeta)$ called Isgur-Wise function \cite{13}. In the case 
of $1^- \to 0^-$ transitions these relations look like
        \begin{eqnarray} &
A_1=\frac{\sqrt{M_PM_V}}{M_P+M_V}(1+\zeta)\xi(\zeta)\; , 
& \nonumber \\ &
A_2=V=\frac{1}{2} \sqrt{\frac{M_P}{M_V}}(1+\frac{M_V}{M_P}) \xi(\zeta) 
\; , & 
        \end{eqnarray}
\noindent
where                    
     $$\zeta=v_P\cdot v_V=\frac{M^2_P+M^2_V-q^2}{2M_VM_P}.$$

Again some dynamical model of mesons is needed to calculate the 
Isgur-Wise function $\xi(\zeta)$ (as an example of such calculations 
see \cite{14,15}). But one can use instead some phenomenologically 
successful parameterization. In particular, the following 
parameterizations was shown \cite{16} to fit experimental data 
reasonably well
        \begin{eqnarray} &
\xi(\zeta)=1-\rho^2(\zeta-1),\;\; \rho \approx 1.08 ; & \nonumber \\ &
\xi(\zeta)=\frac{2}{1+\zeta} \exp{\left\{-(2\rho^2-1)\frac{\zeta-1}
{\zeta+1}\right \} },\;\; 
\rho \approx 1.52; & \nonumber \\ &
\xi(\zeta)=(\frac{2}{\zeta+1})^{2 \rho^2},\;\; \rho \approx 1.45 ; & \\ 
& \xi(\zeta)=\exp{\{-\rho^2(\zeta-1)\}},\;\; \rho \approx 1.37 \; . 
& \nonumber
        \end{eqnarray}

In our case, heavy quark limit can be applied to $\Upsilon \to B^+_c e^- 
\bar \nu_e$ decay. Despite of different analytical forms of the Isgur-Wise 
function, all four parameterizations from (39) give essentially the same 
$Br(\Upsilon \to B^+_c e^- \bar \nu_e)$: $\; 4.1\cdot 10^{-10},\;
3.7\cdot 10^{-10},\; 3.8\cdot 10^{-10}$ and $3.8\cdot 10^{-10}$ 
respectively.

For heavy-light transitions, as for example in $J/ \Psi \to D^-_d e^+ 
\nu_e$ decay, the Isgur-Wise scaling (38) is not applicable. Recently 
B.~Stech proposed \cite{17} a phenomenological model for semileptonic 
form factors which generalizes the Isgur-Wise scaling. It is supposed 
that instead of (38) the following relations hold
       \begin{eqnarray} &
A_1=\frac{\sqrt{M_PM_V}}{M_P+M_V}(1+\zeta) h_{A_1}(\zeta) 
\xi_{PV}(\zeta), \; \;
A_2=\frac{1}{2} \sqrt{\frac{M_P}{M_V}}(1+\frac{M_V}{M_P}) h_{A_2}(\zeta)
\xi_{PV}(\zeta), & \nonumber \\ &
V=\frac{1}{2} \sqrt{\frac{M_P}{M_V}}(1+\frac{M_V}{M_P}) h_V(\zeta) 
\xi_{PV}(\zeta). &
        \end{eqnarray}

The function $\xi_{PV}(\zeta)$ is the same for all form factors for 
the given initial and final states. It approaches the Isgur-Wise 
function in the heavy quark limit.
On contrary, $h$-functions are different for each form factor and 
approach unit in the heavy quark limit.  The concrete expressions for the  
$\xi_{PV}(\zeta)$ and $h$-functions can be found in the original paper 
\cite{17}.

\section{Numerical results} 
To perform numerical calculations within the ISGW model framework, one 
needs to specify quark masses and $\beta$ variational parameters. We use 
the following values for quark masses \cite{10}
$$m_u=m_d=0.33 GeV,\; m_s=0.55 GeV,\; m_c=1.82 GeV,\; m_b=5.12 GeV,$$ 
\noindent
and $\beta$ parameters (in GeV)
       \begin{eqnarray} & 
\beta_k=0.44,\;\beta_{D_d}=0.45,\;\beta_{D_s}=0.56,\;\beta_{B_u}=0.43\; ,
& \nonumber \\ &
\beta_{B_c}=0.92,\; \beta_{\varphi}=0.37,\;  \beta_{J/\psi}=0.62,\;  
\beta_{\Upsilon}=1.1 \; . &
        \end{eqnarray}

All but the last values in (41) are from Table A2 of ref.[10]. The value 
for $\Upsilon$ was obtained by minimizing $<\frac{\vec p^2}{m_b}+V>$, 
with (14) as a trial function and
$V(r)=-\frac{4\alpha_s}{3r}+C+br$, where \cite{10} 
$\alpha_s \approx0.3,\;b=0.18 GeV^2,\; C=-0.84 GeV$.
This minimization problem leads to a cubic equation 
$$\beta^3 - \frac{8\alpha_s m_b}{9 \sqrt{\pi}}\beta^2 - 
\frac{2bm_b}{3 \sqrt{\pi}}=0$$ \noindent
with $\beta \approx 1.1$ as a solution.

Note that this variational solution corresponds to the $\Upsilon$-meson 
mass $M_{\Upsilon}=2m_b + \linebreak <\frac{\vec p^2}{m_b}+V>\approx 
9.44 GeV$ , which should  be compared to the experimental value 
\cite{18} 9.46 GeV.

As was already mentioned, the ISGW model predictions about the high 
$(t_m-t)$-behavior of form factors are not reliable. In numerical 
calculations we use more realistic behavior, suggested in \cite{10} 
(although we don't use other refinements of the model given in \cite{10})
       \begin{equation}
F(t) \to \left(\frac{\beta_P \beta_V}{\beta^2_{PV}}\right)^{3/2}
\left[1+\frac{1}{12}r^2 (t_m-t)\right]^{-2} \; ,
        \end{equation}
\noindent
with
       \begin{equation}
   r^2=\frac{3}{4 m_q m_Q}+\frac{3m^2_Q}{2M_PM_V\beta^2_{PV}}+
   \frac{\Delta r^2}{M_PM_V} .
        \end{equation}

The last term in (43) differs from zero only for $b \to c$ transitions 
and equals \cite{10}  $\Delta r^2 \approx 0.39$.

For the BSW model one needs in addition $1^-$ and $1^+$ pole masses to 
define the form factors $q^2$-dependence. We use the following values 
\newpage
\begin{table}[btph]
\begin{center}
\caption{}
\label{ta:1}
\begin{tabular}{|c|c|c|c|c|c|} \cline{1-6}    \hline
Decay   &$\varphi\to K^+e^-\bar\nu $ & $J/\psi \to D^-_d e^+ \nu$ & 
$J/\psi \to D^-_s e^+ \nu $ & $\Upsilon \to B^+_u e^-\bar \nu $ & 
$\Upsilon \to B^+_c e^-\bar \nu $ \\ \hline
$M_{1^+}, GeV$ & $1.273 (K_1) $&$2.422 (D_1)$&$2.535 (D_{s1})$&$5.745    
$&$6.717$    \\ \hline

$M_{1^-}, GeV$ & $0.892 (K^*) $&$2.010 (D^*)$&$2.112 (D^*_s) $&$5.325 
(B^*)$&$6.317$    \\ \hline
\end{tabular}
\end{center}
\end{table} 

beauty-charm mesons are not yet discovered experimentally. Predictions 
for their masses were taken from \cite{19}
(in particular, $M_{B_c}=6.253 GeV$). Value of $M_{1^+}=5.745 GeV$ 
for $(b \bar u)$-meson is also a potential model prediction taken from 
\cite{20}.

As was already mentioned earlier, we consider $\omega$-parameter of
the BSW model in (28) to be the same as the corresponding 
$\beta$-parameter of the ISGW model from (14). For the 
$\Upsilon \to B^-_c e^-\bar \nu $ decay this choice gives 5-times smaller
branching than it is expected from the heavy quark limit. Especially
sensitive to this parameter is Br($\Upsilon \to B^-_u e^-\bar \nu $),
which is in fact determined by the overlap of the wave function tails,
and it is hard to expect that this tails are correctly given by the 
simple parameterization used in the BSW model. So we decided that it is
more reasonable to choose $\omega_\Upsilon$ such that the heavy quark
limit prediction is reproduced, as much as it is possible, for the 
Br($\Upsilon \to B^-_c e^-\bar \nu $). This gives $\omega_\Upsilon
\approx 2.2GeV$ as compared to $\beta_\Upsilon \approx 1.1GeV$ of the
ISGW model. For other quarkonia $\omega=\beta$ prescription was used.

The numerical results for various semileptonic branching ratios are
summarized in the table below.
\begin{table}[btph]
\begin{center}
\caption{}
\label{ta:2}
\begin{tabular}{|c|c|c|c|c|c|} \cline{1-6}    \hline
Decay   &$\varphi\to K^+e^-\bar\nu $ & $J/\psi \to D^-_d e^+ \nu$ & 
$J/\psi \to D^-_s e^+ \nu $ & $\Upsilon \to B^+_u e^-\bar \nu $ & 
$\Upsilon \to B^+_c e^-\bar \nu $ \\ \hline
ISGW [9] & $7.9\cdot 10^{-15} $&$2.3\cdot 10^{-11}$&$4.8\cdot 
10^{-10} $&$2.9\cdot 10^{-13} $&$1.6\cdot 10^{-10}$    \\ \hline
BSW [8] & $3.1\cdot 10^{-14} $&$3.9\cdot 10^{-11}$&$8.9\cdot 
10^{-10} $&$3.5\cdot 10^{-13} $&$2.0\cdot 10^{-10}$    \\ \hline
Stech [17] & -- &$3.1\cdot 10^{-11}$&$5.2\cdot 
10^{-10} $&$3.0\cdot 10^{-12} $&$3.1\cdot 10^{-10}$    \\ \hline
\end{tabular}
\end{center}
\end{table} 

\section{Conclusions}
We have considered some semileptonic weak decays of vector mesons,
using the most popular ISGW and BSW quark models. The predictions
of these models agree to each other reasonably well (within a factor 2),
except $\varphi\to K^+e^-\bar\nu $ decay, where predicted branchings
differ 4-times.

The corresponding branching ratios were also calculated using recently
proposed Stech's phenomenological model \cite{17}. The results agree
again with the ISGW and BSW models predictions, except $\varphi\to 
K^+e^-\bar\nu $ and $\Upsilon \to B^+_u e^-\bar \nu $ decays.
As for the $\varphi\to K^+e^-\bar\nu $ decay, for which the result is
$Br(\varphi\to K^+e^-\bar\nu)=2.7 \cdot 10^{-12}$, we don't  expect
Stech's model to be valid for it. But it is interesting to note that
if we don't require, as in \cite{17}, $\xi_{PV}(\zeta)$ to have a pole
in $q^2$ at the position of the lowest $0^-$ resonance (the pseudoscalar
$P$ meson itself), but instead demand that the pole position for 
$\xi_{PV}(\zeta)$ depends on the form-factor, in which $\xi_{PV}(\zeta)$
enters, exactly as in the BSW model (that is $1^-$-pole for the $V$
form-factor and $1^+$-pole for the $A_1$ and $A_2$ form-factors), then
so modified Stech's model predicts  $Br(\varphi\to K^+e^-\bar\nu)=
9.0 \cdot 10^{-15}$, again close to the ISGW and BSW results. The other
decay modes are not significantly effected by this modification. In
particular, an order of magnitude difference between Stech's model on
one hand and ISGW or BSW model on another for the $\Upsilon \to 
B^+_u e^-\bar \nu $ decay still persists. It seems to us that the Stech's
model has difficulties in handling this decay mode.

Unfortunately, the predicted branching ratios are too small and so
an experimental study of the decays considered is questionable in 
near future.

\section*{Acknowledgments}
We are grateful to Victor Chernyak for useful discussions.


\newpage
\begin{thebibliography}{99}
\bibitem{1} A.~Le Yaouanc, Nucl. Instr. Meth. {\bf A351}(1994), 15.
\bibitem{2} N.~Isgur, Models of semileptonic decays, Invited talk given
at 1989 Int. Conf. on Heavy Quark Physics, Cornell. Toronto University
preprint UTPT-89-25, 1989.
\bibitem{3} M.~Wirbel, Semileptonic B decays, Dortmund University preprint
DO-TH-89/4, 1989
\bibitem{4} D.~Melikhov, Phys. Rev. {\bf D53(1996)}, 2460.
\bibitem{5} G.~Altarelli et al., Nucl. Phys. {\bf B208}(1982), 365.
\newline  N.~Cabibbo, G.~Corbo, L.~Maiani, 
Nucl. Phys. {\bf B155}(1979), 93.  
\bibitem{6} B.~Grinstein, M.~B.~Wise, N.~Isgur, Phys. Rev. Lett. {\bf 56}
(1986), 298.
\bibitem{7} D.~Scora, N.~Isgur, Phys. Rev. {\bf D40(1989)}, 1491.
\bibitem{8} M.~Wirbel, B.~Stech, M.~Bauer, Z. Phys. {\bf C29}(1985), 637.
\newline    M.~Bauer, M.~Wirbel, Z. Phys. {\bf C42}(1989), 671.
\bibitem{9} N.~Isgur, D.~Scora, B.~Grinstein, M.~B.~Wise,
Phys. Rev. {\bf D39(1989)}, 799.
\bibitem{10} D.~Scora, N.~Isgur, Phys. Rev. {\bf D52(1995)}, 2783.
\bibitem{11} T.~Altomari, L.~Wolfenstein, Phys. Rev. {\bf D37(1988)}, 681.
\bibitem{12} B.~Konig, J.~G.~Korner, M.~Kramer, P.~Kroll, 
Infinite Momentum Frame calculation of semileptonic heavy $\Lambda_b \to
\Lambda_c$ transitions including HQET improvements, preprint 
DESY-93-011, 1993 (hep-ph/ 9701212).
\bibitem{13} N.~Isgur, M.~B.~Wise, Phys. Lett. {\bf B232}(1989), 113; 
{\bf B237}(1990), 527.
\newline M.~Neubert, V.~Rieckert, Nucl. Phys. {\bf B382}(1992), 97.
\bibitem{14} A.~Le Yaouanc, L.~Oliver, O.~P\'{e}ne, J.~-C.~Raynal
Phys. Lett. {\bf B365}(1996), 319. 
\bibitem{15} D.~Melikhov, Heavy quark expansion and universal form-factors
in quark model, hep-ph/ 9706417.
\bibitem{16} H.~Albrecht et al. (ARGUS coll.), 
Z. Phys. {\bf C57}(1993), 533.
\bibitem{17} B.~Stech, Z. Phys. {\bf C75}(1997), 245; Nucl. Phys. Proc. Suppl.
{\bf 50}(1996), 45.
\bibitem{18} Review of Particle Physics, Phys. Rev. {\bf D54}(1996).
\bibitem{19} S.~S.~Gershtein, V.~V.~Kiselev, A.~K.~Likhoded,
A.~V.~Tkabladze, Usp. Fiz. Nauk {\bf 165}(1995), 3. 
\bibitem{20} S.~N.~Gupta, J.~M.~Johnson, Phys. Rev. {\bf D51}(1995), 168.

\end{thebibliography}
\end{document}